\documentclass[%
reprint,
amsmath,amssymb,
aps,
floatfix,
]{revtex4-1}
\usepackage{graphicx}
\usepackage{bm}
\usepackage{hyperref}
\usepackage{csquotes}
\usepackage{amsmath}
\usepackage{caption}
\usepackage{subcaption}
\begin{document}
\preprint{APS/123-QED}
\title{Effect of tensor force on lowering of $5/2_{1}^{-}$ level in heavier Cu isotopes}
\author{Kanhaiya Jha}
\author{P. K. Raina}
\affiliation{Indian Institute of Technology Ropar, Rupnagar, Punjab-140001, INDIA}
\date{\today}
\begin{abstract}
The inversion of $3/2^{-}$ and $5/2^{-}$ levels in heavier Cu isotopes is one of the most visible example of shell-evolution, caused by the strong monopole attraction between the nucleons occupying in orbitals $\pi 0f_{5/2}$ and $\nu 0g_{9/2}$.  The tensor part of the nucleon-nucleon interaction is expected to be the driving force behind the monopole migration. In shell model framework, usually spin-tensor decomposition is used to get the information of individual force components to the shell evolution, however, in the present scenario, this method can not apply on \textit{pfg} model space due to the missing spin-orbit partners $ 0f_{7/2}$ and $ 0g_{7/2}$. Therefore, we have analytically obtained the tensor force two-body matrix elements (TBME) for \textit{pfg} model space using Yukawa potential, and subtract it from effective interaction jj44b. The interaction without tensor part, named as jj44a, have been used for calculation of Ni, Zn, Ge and Cu isotopes with various physics viewpoints. In most of the cases, the theoretical results are in good agreement with the experiment only when tensor force is included to the interaction jj44a. 

\end{abstract}
\pacs{Valid PACS appear here}
\maketitle
\section{Introduction}
\label{sec1}
Over the years, many new predications and observations in nuclear physics is possible due to the progress in sophisticated experimental facilities, availability of radioactive ion beams, and enhanced computational power across the globe. The studies reveal that the well established magic numbers are not permanent, but subject to unbalanced \textit{N/Z} ratio coupled with the unique characteristics of individual force components \cite{caurier,brown, sorlin, otsukare}. The sudden drop of the $5/2_{1}^{-}$ state in heavier Cu isotopes is one of the most visible example of shell evolution around the double magic nuclei $^{78}$Ni \cite{sieja, sahin,flanagan,ichikawa}. The shift of $5/2_{1}^{-}$ level in heavier Cu isotopes refers the strong attractive monopole interaction between protons and neutrons occupying in $\pi 1f_{5/2}$ and $\nu 0g_{9/2}$ orbitals, respectively \cite{franchoo,otsuka01, otsuka05,smirnova1}. The monopole Hamiltonian has large influence on the large-scale shell model calculations because of the significant role at shell structure evolution

In general, the shell evolution is  generated by monopole components of the effective NN interactions, defined as  \cite{otsukare}
\begin{equation}
\bar{V}^{T}_{jj'} = \frac{\Sigma_{J} (2J+1) <jj'| V |jj'>_{JT}}{\Sigma_{J} (2J+1)},
\end{equation}
where $\bar{V}^{T}_{jj'}$ is J-averaged two-body matrix elements (TBMEs) corresponds to the Isospin T and orbitals $j$ and $j'$. 

In recent years, the tensor force monopole matrix elements ($\bar{V}^{T}_{jj'}(\zeta)$) gains lot of interest due to its unique and robust features \cite{otsuka05, sunoda, umeya}.  It shows opposite sign for the configurations $j_{>}j'_{>}$ and $j_{>}j'_{<}$, where $j_{>} = l + s$ and $j_{<} = l - s$ represents spin-up and spin-down orbitals, respectively. For example, the tensor force interaction between orbitals $\pi 0f_{5/2}$ and $\nu 0g_{9/2}$ ($j_{<}-j'_{>}$) is attractive whereas repulsive for $\pi 0f_{7/2}$ ($j_{>}$) and $\nu 0g_{9/2}$ ($j'_{>}$), results in evolution of Z = 28 shell gap for 40 $<$ N $<$ 50. Further, it has been also reported that the tensor components of the NN interaction drives the monopole migration of $5/2_{1}^{-}$ state, whereas mechanism responsible for the steep lowering of $1/2_{1}^{-}$ state is still not clear. In the present work, our main goal is to understand the shell evolution in $pf_{5/2}g_{9/2}$ model space due to the tensor force components of the NN interaction. Since the spin-tensor decomposition (STD) \cite{kirson,klingenback,kenji}  method which is used to extract the strength of individual force components of the shell-model effective interactions can only apply to the model space containing all spin-orbit partners \cite{smirnovaplb}, for eg. \textit{p}, \textit{sd}, \textit{psd}, \textit{pf}, \textit{sdpf}-shell. On the other hand, in the $pf_{5/2}g_{9/2}$ model space, spin-orbit partners $ 0f_{7/2}$ and $ 0g_{7/2}$ are missing. In our recent work \cite{kjha,kjha1}, we have used analytical expression of tensor force to improve the disparities present in \textit{p} \cite{cohen} and \textit{pf}-shell \cite{honma2} effective interactions, and we get rather good result as expected. In these studies, we found that the calculated tensor force monopole matrix elements retains systematic trends originating in the bare tensor force. Therefore, we have applied a similar approach in \textit{pfg} model space. Here, we also find that the calculated tensor force monopole matrix elements possess systematic features of tensor force in both Isospin T = 0 and 1, discussed later in the article in detail.

The jun45\cite{honma} and jj44b\cite{jj44b} are commonly used effective interactions for \textit{$pf_{5/2}g_{9/2}$} model space consisting of single-particle orbitals $1p_{3/2}$,  $ 0f_{5/2}$, $ 1p_{1/2}$, and $ 0g_{9/2}$ on top of $^{56}$Ni core. Since the excitation spectra of Ni and Cu isotopes  have been included for derivation of interaction jj44b, therefore, the interaction jj44b give rather reasonable results than the jun45 for heavier Cu isotopes, particularly for $5/2_{1}^{-}$ and $1/2_{1}^{-}$ states \cite{vingerhoets}. The Ni and Cu isotopes are potential candidates of our study, hence, interaction jj44b is used in the present work.  In this manuscript, our main aim is to show the importance of tensor force, therefore, we have subtracted tensor part
from the interaction jj44b, and named as interaction jj44a. The interactions with (jj44b) and without (jj44a) tensor part are then used to calculate proton effective single-particle energy of $pf_{5/2}g_{9/2}$-orbitals in Cu isotopes, 2$^{+}$ excitation energy of even-even Ni, Zn and Ge isotopes, low-lying yrast level structure of Cu isotopes and  electromagnetic properties of Ni, Cu, Zn and Ge isotopes.  The effect of tensor force 
can be seen clearly understood by observing changes in the results of interactions jj44a and jj44b.

The present paper is arranged as follows: Sec.~\ref{sec2} discussed about theoretical framework to calculate tensor force two-body matrix elements for \textit{pfg}-shell. In Sec.~\ref{sec3} of the results and discussion, the Cu, Ni, Zn and Ge isotopes have been presented with various physics viewpoints. The summary of this work is finally shown in Sec.~\ref{sec4}.

\begin{figure*}
	\centering
	\resizebox{0.9\textwidth}{!}{%
		\includegraphics{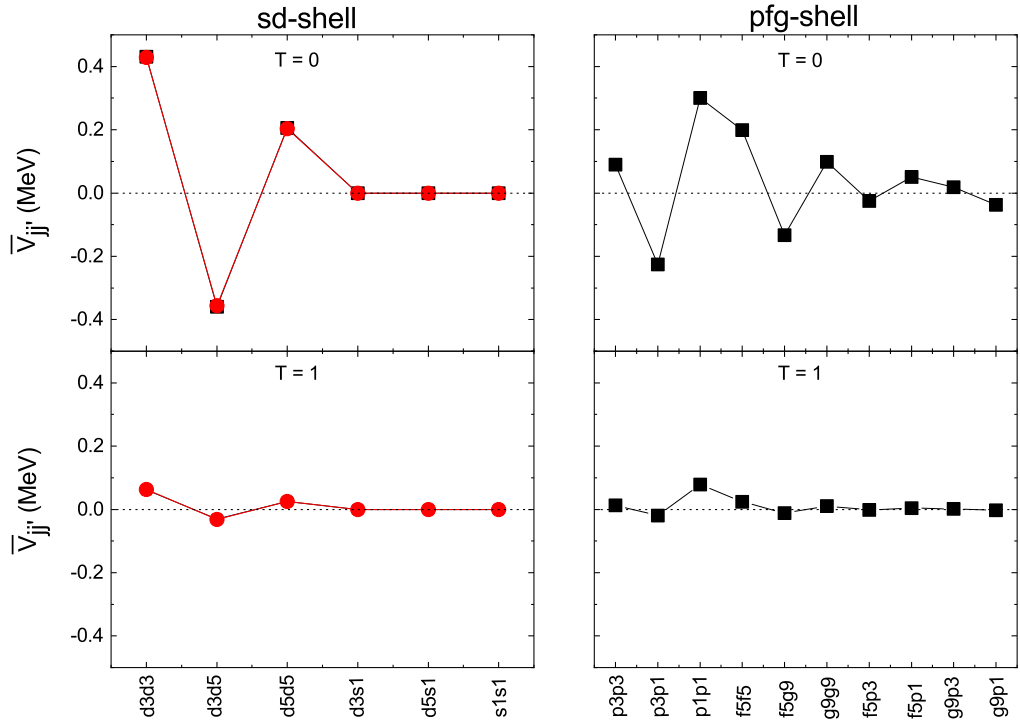}
	}
	\vspace{-1.0cm}
	\caption{(Color online) Calculated (solid square) tensor force monopole matrix elements in \textit{sd}-shell and\textit{ pfg}-shell along with the those of USDB (solid circle) for Isospin T = 0 and 1 (see text for details). } 
	\label{F1}      
\end{figure*}
\begin{figure}
	\centering
	\resizebox{0.50\textwidth}{!}{%
		\includegraphics{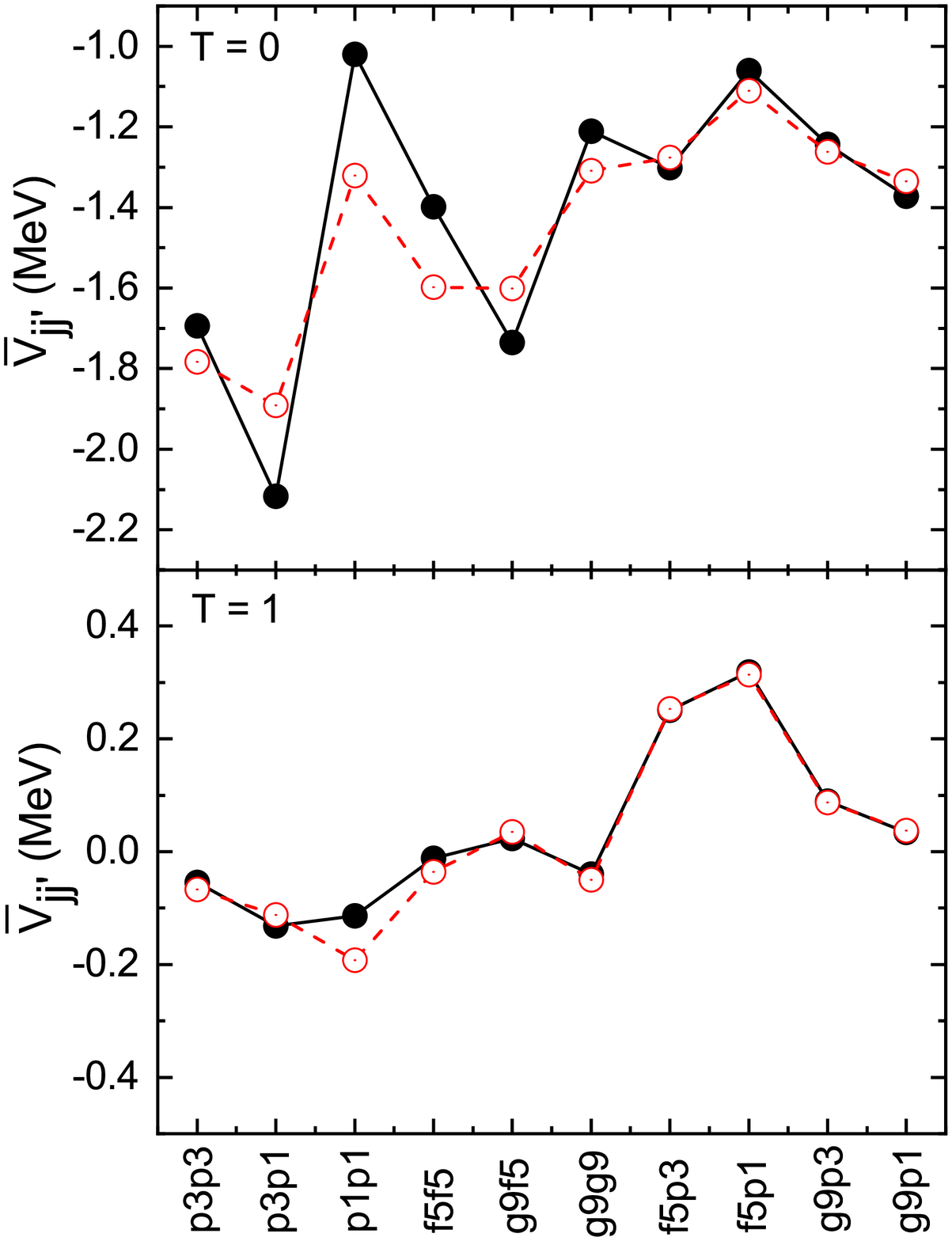}
	}
	\vspace{-1.4cm}
	\caption{(Color online) J-averaged two-body matrix elements of interactions jj44b and jj44a. The solid (open) symbol is used for interaction jj44b (jj44a).} 
	\label{F2}      
\end{figure}
\section{Theoretical formalism}
\label{sec2}
The interaction between two-nucleon is defined as the linear sum of the scalar product of configuration space operator \textit{Q} and spin space operator \textit{S} of rank k:
\begin{equation}
	V= \sum_{k = 0}^{2} V(k) = \sum_{k = 0}^{2} Q^{k} .  S^{k} , 
\end{equation}

Based on the tensor structure, the shell-model effective interaction is decomposed to the individual components as\\

$<ab: JT|V(k)|cd: JT> ~ =$
\[
\begin{aligned}
~ = \frac{1}{\sqrt{(1+\delta_{n_{a}n_{b} l_{a} l_{b} j_{a} j_{b}}) (1+\delta_{n_{c}n_{d} l_{c} l_{d} j_{c} j_{d}})}} \\	
\sum_{LSL'S'} \
\sqrt{(2j_{a}+1) (2j_{b}+1)(2L+1)(2S+1)}
\begin{Bmatrix}
l_{a} & 1/2 & j_{a}\\
l_{b} & 1/2 & j_{b}\\
L & S & J
\end{Bmatrix}\\
\sqrt{(2j_{c}+1) (2j_{d}+1)(2L'+1)(2S'+1)}
\begin{Bmatrix}
l_{c} & 1/2 & j_{c}\\
l_{d} & 1/2 & j_{d}\\
L' & S' & J
\end{Bmatrix}\\
(2k+1) (-1)^{J}
\begin{Bmatrix}
L & S & J\\
S' & L' & k
\end{Bmatrix}
\sum_{J'}(-1)^{J'} (2J'+1)
\begin{Bmatrix}
L & S & J'\\
S' & L' & k
\end{Bmatrix}\\
\end{aligned}
\]
\[
\begin{aligned}
\sum_{j'_{a} j'_{b}j'_{c} j'_{d}}
\sqrt{(2j'_{a}+1) (2j'_{b}+1)(2L+1)(2S+1)}
\begin{Bmatrix}
l_{a} & 1/2 & j'_{a}\\
l_{b} & 1/2 & j'_{b}\\
L & S & J'
\end{Bmatrix}\\  
\sqrt{(2j'_{c}+1) (2j'_{d}+1)(2L'+1)(2S'+1)}
\begin{Bmatrix}
l_{c} & 1/2 & j'_{c}\\
l_{d} & 1/2 & j'_{d}\\
L' & S' & J'
\end{Bmatrix}\\  
\sqrt{(1+\delta_{n_{a}n_{b} l_{a} l_{b} j'_{a} j'_{b}}) (1+\delta_{n_{c}n_{d} l_{c} l_{d} j'_{c} j'_{d}})}\\
<\alpha \beta: JT|V|\gamma \delta : JT>\\
\end{aligned}
\]
where, rank k = 0, 1, and 2 represent central, spin-orbit, and
tensor forces, respectively. The $a$ = $(n_{a} l_{a} j_{a})$, $\alpha$ = $(n_{a} l_{a} j'_{a})$ are shorthand notation for the set of quantum numbers.\\

Despite the success of STD method for \textit{p}, \textit{sd}, \textit{psd}, \textit{pf} and \textit{sdpf} model space, this method has certain limitations.
The method can not be apply on the model space where spin-orbit partners within the oscillator shell are missing \cite{smirnovaplb}. For example, missing spin-orbit partners $0f_{7/2}$ and $ 0g_{7/2}$ in $pf_{5/2}g_{9/2}$ model space. This mathematical restriction is due to the presence of summation over j terms in the right-side of the final expression of STD obtained from the \textit{jj}-coupled matrix elements. The expression require both j orbitals i.e. $l+\frac{1}{2}$ and $l-\frac{1}{2}$ associated with orbital quantum number \textit{l} to decompose an effective interaction. 

We have analytically calculated the tensor force matrix elements using expression
\begin{equation}
	V_{\zeta}= V(r)\sqrt{\frac{24 \pi}{5}}[Y^{(2)}.{(\sigma_{1} X  \sigma_{2} )}^{(2)}](\tau_{1} . \tau_{2}), 
\end{equation}
where radial dependency is treated with the Yukawa potential
\begin{equation}
	V(r) = -V_{0} \frac{e^{-r/a}}{r/a}, 
\end{equation} 
where Compton scattering length of pion ``\textit{a}'' is 1.41 fm for $m_{\pi}$ = 139.4 MeV and strength parameter $V_{0}$, in our calculation, is obtained from the fit of $\bar{V}^{T=0}_{jj'} (\zeta)$ matrix elements of USDB \cite{brown2} obtained using spin-tensor decomposition (STD) method. 
 
On the left side panel of Fig.~\ref{F1}, the $\bar{V}^{T}_{jj'} (\zeta)$ calculated and USDB are shown for both T = 0 and 1. The analytically calculated tensor force matrix elements completely reproduce the tensor force matrix elements of USDB. Further, we have recently used this method to improve the tensor force disparities present in the effective interactions CK (8-16) and GX1B1 of  \textit{p}-shell and \textit{pf}-shell, respectively \cite{kjha,kjha1}. In these studies, we find that the replacement of tensor force TBMEs with calculated ones significantly improve the theoretical results. On the same line in the present work, we have used this approach for \textit{pfg} model space. As  seen from right-side panel of Fig.~\ref{F1}, we find that the calculated tensor force monopole matrix elements possess systematic features of tensor force in both Isospin T = 0 and 1, and the strength of $\bar{V}^{T = 0}_{jj'} (\zeta)$ is greater than that of $\bar{V}^{T = 1}_{jj'} (\zeta)$ as expected. Further, the calculated tensor force TBMEs are subtracted from the corresponding TBMEs of interaction jj44b. The interaction without tensor part is named as jj44a. In Fig.~\ref{F2}, the monopole matrix elements of interactions with (jj44b) and without (jj44a) tensor force are shown for both T = 0 and 1.  The deviations from solid lines are due to the tensor force.

\section{Result and Discussion}
\label{sec3}
\subsection{Effective single-particle energy}
\label{sec3a}
The effective single-particle energy (ESPE) is a good probe to test the 
sensitivity of TBMEs and change in the single-particle energy orbitals. The ESPE is corresponds to the spherical shell structure which is calculated as separation energy of a nucleon added or removed from the closed shell configuration. The expression of the effective single-particle energy is given as \cite{smirnova1}
\begin{equation}
\epsilon^{'\rho'}_{j^{'}}(A)= \epsilon^{\rho'}_{j^{'}}+ \sum_{j} \hat{n}_{j}^{\rho}  \bar{V}_{jj'}^{\rho \rho'}(A),
\end{equation}
where $\bar{V}_{jj'}^{\rho \rho'}(A)$ is mass dependent monopole matrix elements, and $\hat{n}^{\rho}_{j}$ is the number of neutron in the valence orbital j.

\begin{figure}
	\centering
	\resizebox{0.50\textwidth}{!}{%
		\includegraphics{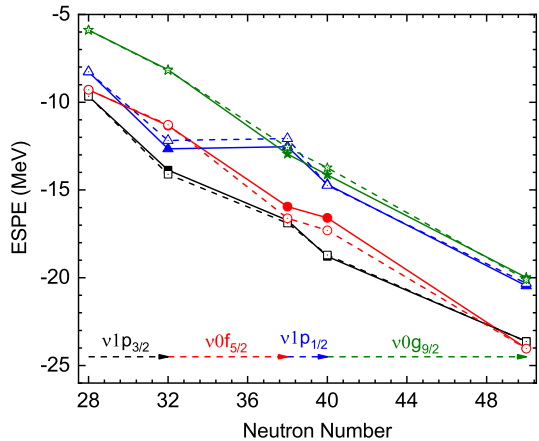}
	}
	\vspace{-0.7cm}
	\caption{(Color online) Evolution of proton ESPE of $pfg-$orbitals with neutron number. The proton orbitals $p_{3/2}$, $f_{5/2}$, $p_{1/2}$ and
	$g_{9/2}$ are represented by solid (open) square, circle, triangle, and star for interaction jj44b (jj44a), respectively.} 
	\label{F3}      
\end{figure}

In Fig.~\ref{F3}, the ESPEs of proton orbitals in Cu isotopes are shown with interactions jj44a and jj44b. The deviations of the single-particle energy orbitals from solid lines are due to the tensor force. The Cu (Z = 29) isotopes having one proton on top of $^{56}$Ni (Closed shell) is one of the best choices to analyze the monopole contribution as particle-particle correlations are often negligible in case of $CS \pm 1$ nuclei. For $^{57}$Cu, the $ \epsilon_{j'}^{' \pi} $ is equal to the their unperturbed single-particle energies $ \epsilon_{j'}^{ \pi} $ as there is no neutrons available in the valence orbitals. As the neutrons start filling in the valence orbitals,  the ESPEs of proton orbitals goes downward except $\pi 1p_{1/2}$ orbital for 32 $<$ N $<$ 38. The downward shift of proton single-particle energy orbitals is mainly due to the attractive T = 0 monopole matrix elements. The spin-orbit partners gap $\pi (1p_{1/2}-1p_{3/2})$ increases with the filling of $\nu 0f_{5/2}$ orbital whereas decreases by almost $\sim$ 1.0 MeV with the filling of $\nu 0g_{9/2}$ orbital. Since the 
tensor force interaction between $\pi 1p_{3/2}$ and $\nu 0f_{5/2}$ is attractive, whereas repulsive between $\pi 1p_{1/2}$ and $\nu 0f_{5/2}$. Hence, tensor force acts towards lowering of spin-orbit partners gap $\pi (1p_{1/2}-1p_{3/2})$ from N = 32 to 38. Likewise, the tensor force reduces this gap when $\nu 0g_{9/2}$ orbital is filled. The SPE shift of $\pi 1p_{1/2}$ orbital from the solid line is more pronounced for 28 $<$ N $<$ 38 is mainly due to the large attractive tensor force $\bar{V}_{13}^{T = 0}$. Further, it can be clearly seen from Fig.~\ref{F3} that the tensor force has significant contribution in lowering of $\pi 1f_{5/2}$ orbital. The gap $\pi (0f_{5/2}-1p_{3/2})$ is increase and decrease alternatively with the filling of neutrons in the $pf_{5/2}g_{9/2}$ orbitals. The SPE $\pi 0f_{5/2}$ drops significantly with filling of $\nu 0g_{9/2}$ orbital, and cross the $\pi 1p_{3/2}$ around $^{75}$Cu. It is clearly evident from Fig.~\ref{F3} that the slope of $\pi 0f_{5/2}$ orbital is more steeper from $^{69}$Cu to $^{79}$Cu when tensor force is included to the interaction jj44a. The evolution of Z = 28 shell gap between the $\pi 0f_{7/2}$ and $\pi 1p_{3/2}$ orbitals for 40 $<$ N $<$ 50 is also reported due to the repulsive tensor force interaction between orbitals $\pi 0f_{7/2}$ and $\nu 0g_{9/2}$ \cite{sahin}.

\begin{figure*}
	\centering
	\resizebox{1.0\textwidth}{!}{%
		\includegraphics{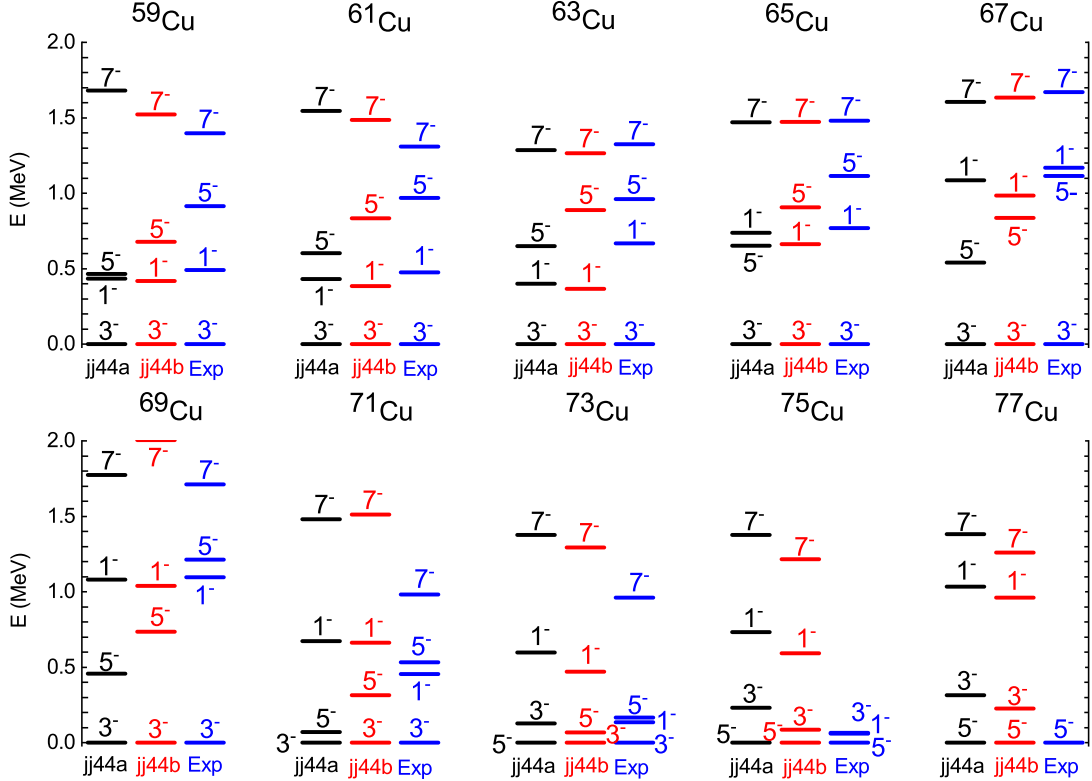}
	}
	\vspace{-1.5cm}
	\caption{(Color online) Level structure of $^{59-77} $Cu isotopes. Theoretical calculation are performed with the interactions jj44a and jj44b, and experimental data are taken from the Refs.\cite{nndc,vingerhoets}.} 
	\label{F4}      
\end{figure*}
\subsection{Excitation spectra of $^{59-77}$Cu }

In Fig.~\ref{F4}, the yrast $3/2_{1}^{-}$, $5/2_{1}^{-}$, $1/2_{1}^{-}$ and $7/2_{1}^{-}$ states of $^{59-77}$Cu are shown with the effective interactions jj44a and jj44b. The theoretical calculations have been performed with shell-model code NUSHELLX@MSU \cite{nushell}, and experimental data are taken from Ref.\cite{nndc}. The comparison of low-lying excitation spectra of $^{57-79}$Cu isotopes with effective interactions jun45 and jj44b are reported in Ref.\cite{vingerhoets}. It has been found that $1/2_{1}^{-}$ and $5/2_{1}^{-}$ states are predicted much higher than the measured value by interaction jun45. On the other hand, jj44b reasonably predicts these states since excitation spectra of Ni and Cu are included while deriving interaction jj44b. Hence, we have used interaction jj44b in the present work. 
 
For $^{57}$Cu since no neutrons are present in the valence orbitals, the excitation spectra of $3/2_{1}^{-}$, $5/2_{1}^{-}$, and $1/2_{1}^{-}$ states are related to their single-particle energies -9.6566 MeV, -9.2859 MeV, and -8.2695 MeV, respectively. For $^{59}$Cu which is only two neutrons away from the $^{57}$Cu, the excitation energies are changed substantially due to the proton-neutron interactions. The $5/2_{1}^{-}$ state of $^{57}$Cu which is in between the states $3/2_{1}^{-}$ and $1/2_{1}^{-}$ has been shifted above $1/2_{1}^{-}$ state in $^{59}$Cu. 
For $^{59}$Cu, the interaction without tensor force predicts $1/2_{1}^{-}$ and $5/2_{1}^{-}$ states very close to each other whereas these states are separated by 0.423 MeV in the experiment. The inclusion of tensor force predicts the desired gap between the $1/2_{1}^{-}$ and $5/2_{1}^{-}$ states. It can be followed from Fig.~\ref{F3} that the single-particle energy orbital $\pi 1p_{1/2}$ falls rapidly than $\pi 0f_{5/2}$ with filling of neutrons in $\nu 1p_{3/2}$. The $5/2_{1}^{-} $ level is observed almost constant between 0.91 MeV and 1.11 MeV for $^{59- 67}$Cu whereas it is 1.21 MeV, 0.53  MeV, and 0.17 MeV for $^{69}$Cu, $^{71}$Cu and $^{73}$Cu, respectively. The calculation predicts $5/2^{-}_{1}$ level above the $1/2^{-}_{1}$ for $^{59-65}$Cu and beyond $^{65}$Cu reverse in order. The plausible reason is the crossing of single-particle energy orbitals $\pi 0f_{5/2}$ and $\pi 1p_{1/2}$ around $^{63}$Cu, shown in Fig.~\ref{F3}. Although, the interaction jj44b predicts $1/2^{-}_{1}$ and $ 5/2^{-}_{1}$ states of
 $^{59-67}$Cu below the measured value but maintains the experimentally measured order, however, the $5/2^{-}_{1}$ and $1/2^{-}_{1}$ states of $^{71,73}$Cu are predicted in reverse order. The sudden drop of $5/2_{1}^{-} $ level from N = 40 to 50 suggested that $5/2_{1}^{-} $ level becomes the ground state in heavier Cu isotopes. The recent studies made by K. T. Flanagan \textit{et.al} \cite{flanagan} based on the observation of magnetic moments and spin of $^{69,71, 73}$Cu confirmed that the $5/2^{-} $ is the ground state spin of $^{75}$Cu. In the literature, it has been suggested that the attractive tensor force interaction between $\pi 0f_{5/2}$ and $\nu 0g_{9/2}$ orbitals may be responsible for the migration of the $5/2_{1}^{-}$ state in heavier Cu isotopes. Likewise $5/2_{1}^{-}$ state, the experimentally measured $1/2_{1}^{-}$ state also comes down significantly from $\approx$ 1.10 MeV for $^{67,69}$Cu to 61 KeV for $^{75}$Cu with the filling of $ \nu 0g_\frac{9}{2}$ orbital, however,  it is still interesting to know whether similar physics mechanism as in case of $5/2_{1}^{-}$ state is responsible for the lowering of $1/2_{1}^{-}$ state. It has been observed that the tensor force lowers $1/2_{1}^{-}$ state throughout the Cu isotopic chain.  The gap between $1/2_{1}^{-}$ and $5/2_{1}^{-}$ states of $^{59-65}$Cu increases by tensor force whereas this gap decreases thereafter. The interaction jj44a predicts ground state spin of $^{73}$Cu as $5/2^{-}$ whereas the measured spin is $3/2^{-}$. The inclusion of tensor force correctly reproduce the ground state spin  of $^{73}$Cu by inverting $3/2^{-}$ and $5/2^{-}$ states. The experimentally measured ground state spin of $^{75}$Cu is $ 5/2^{-}$, which is nicely reproduced by jj44b. As followed from the Fig.~\ref{F4}, inversion of $ 3/2^{-}$ and $ 5/2^{-}$ states from $^{71}$Cu to $^{73}$Cu are predicted by interaction jj44a. The tensor force reasonably reproduced the $ 3/2^{-}$ and $ 5/2^{-}$ gap between them. Interestingly, the gap between $3/2^{-}$ and $1/2^{-}_{1}$ states remains almost unchanged by the tensor force beyond $^{67}$Cu, whereas the gap between $5/2_{1}^{-}$ and $1/2_{1}^{-}$ states are minimized by the tensor force. This is also observed that the single-particle orbitals $\pi 1p_{3/2}$ and $\pi 1p_{1/2}$ are almost unchanged by tensor force from $^{69}$Cu to $^{79}$Cu. The interaction jj44b reasonable predict the drop of $7/2_{1}^{-}$ excitation energy from 1.711 MeV in $^{69}$Cu to 0.981 MeV in $^{71}$Cu. The calculation shows that $7/2_{1}^{-}$ state drops to $\approx$ 1.2 MeV at $^{75}$Cu from the
maximum value $\approx$ 2.0 MeV at $^{69}$Cu. This  state is interpreted as corresponding $\pi p_{3/2}^{1} \otimes 2^{+}$ ($^{68, 70, 72}$Ni). The $7/2^{-}_{1}$ state remains almost unchanged after addition of tensor force for $^{63, 65, 67}$Cu whereas decreases in case of $^{59, 61, 73, 75, 77}$Cu.
\begin{figure}
	\centering
	\resizebox{.5\textwidth}{!}{%
		\includegraphics{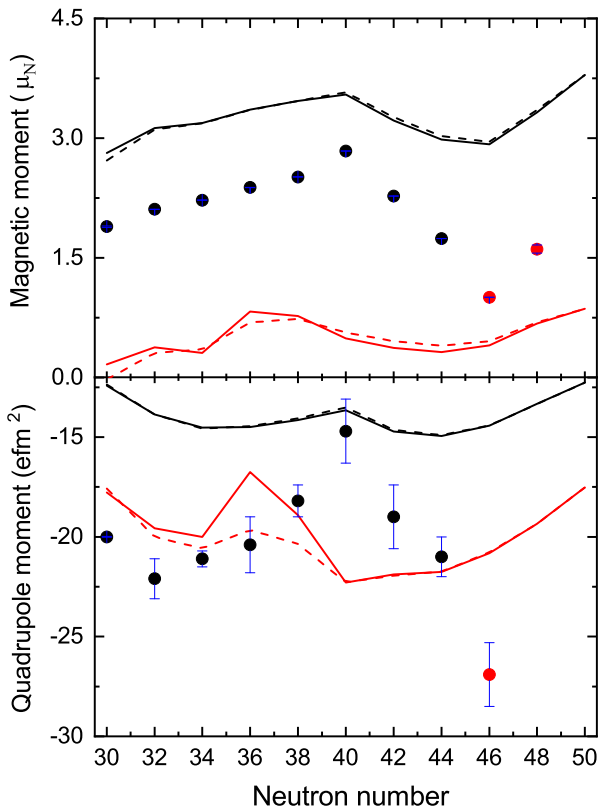}
	}
	\vspace{-1.5cm}
	\caption{(Color online) Theoretical magnetic moments (top) and quadrupole moments (bottom) are calculated using jj44a (dashed lines) and jj44b (solid lines) interactions. The black (red) lines indicates the $3/2^{-}$ ($5/2^{-}$) states, and corresponding experimental data by filled circle.  } 
	\label{F5}      
\end{figure}

\begin{figure*}
	\centering
	\resizebox{1.0\textwidth}{!}{%
		\includegraphics{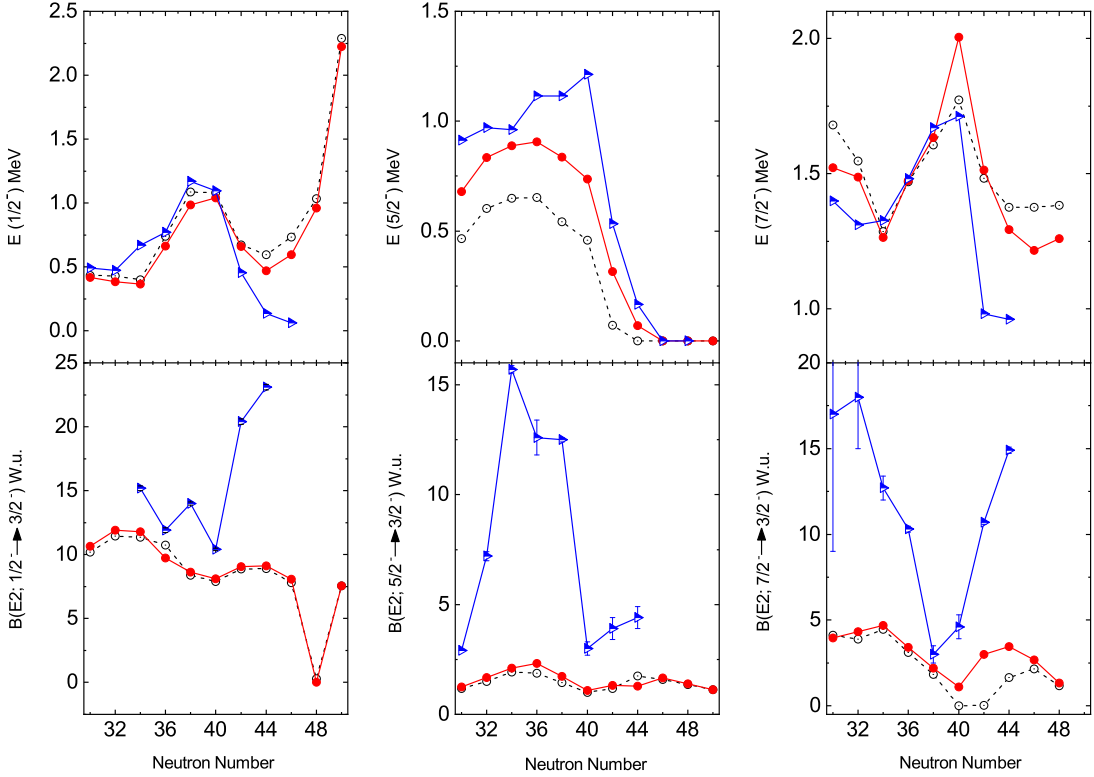}
	}
	\vspace{-1.5cm}
	\caption{(Color online) Top: Excitation energies of $1/2^{-}$, $5/2^{-}$ and $7/2^{-}$ states of $^{69-77}$Cu. Bottom: The B(E2) value of $1/2^{-}$, $5/2^{-}$, and $7/2^{-}$ states of $^{69-79}$Cu. The shell-model calculations are performed with interactions jj44a (open circle) and jj44b (filled circle), and the experimental data (half-filled triangle) is taken from the Ref.\cite{nndc}. The standard value of effective charges $e_{\pi}$ = 1.5e and $e_{\nu}$ = 0.5e were used to calculate the B(E2).} 
	\label{F6}      
\end{figure*}
\begin{figure}
	\centering
	\resizebox{.5\textwidth}{!}{%
		\includegraphics{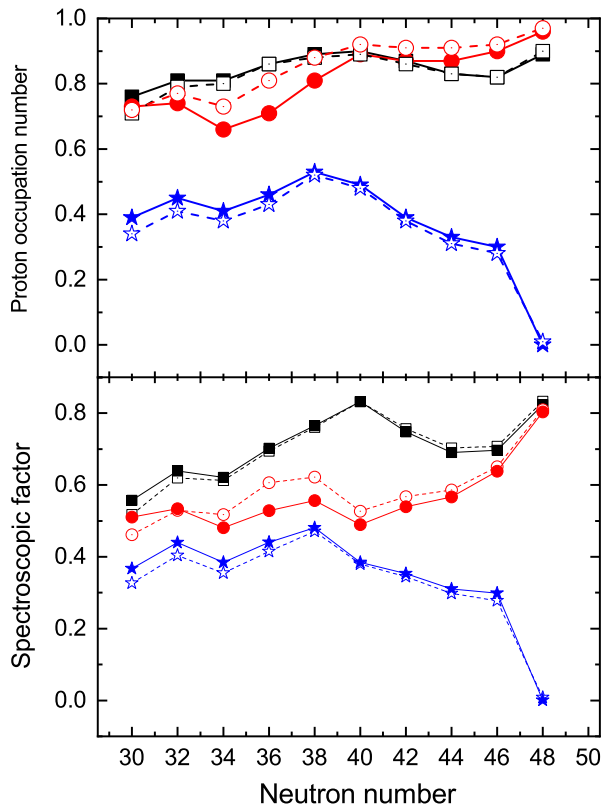}
	}
	\vspace{-1.5cm}
	\caption{(Color online) Top: Proton occupation number of proton orbitals for $^{59-79}$Cu.  Bottom: Spectroscopic factor of $^{A}$Cu calculated with respect to proton transfer to $^{A-1}$Ni. The shell model calculations are performed with interactions jj44a (open symbols) and jj44b (filled symbols). The symbols square, circle and star are representing $3/2^{-}$, $5/2^{-}$ and $1/2^{-}$ states, respectively. } 
	\label{F7}      
\end{figure}
\begin{figure*}
	\begin{subfigure}{0.3\linewidth}
		\centering
		\includegraphics[width=1.30\textwidth]{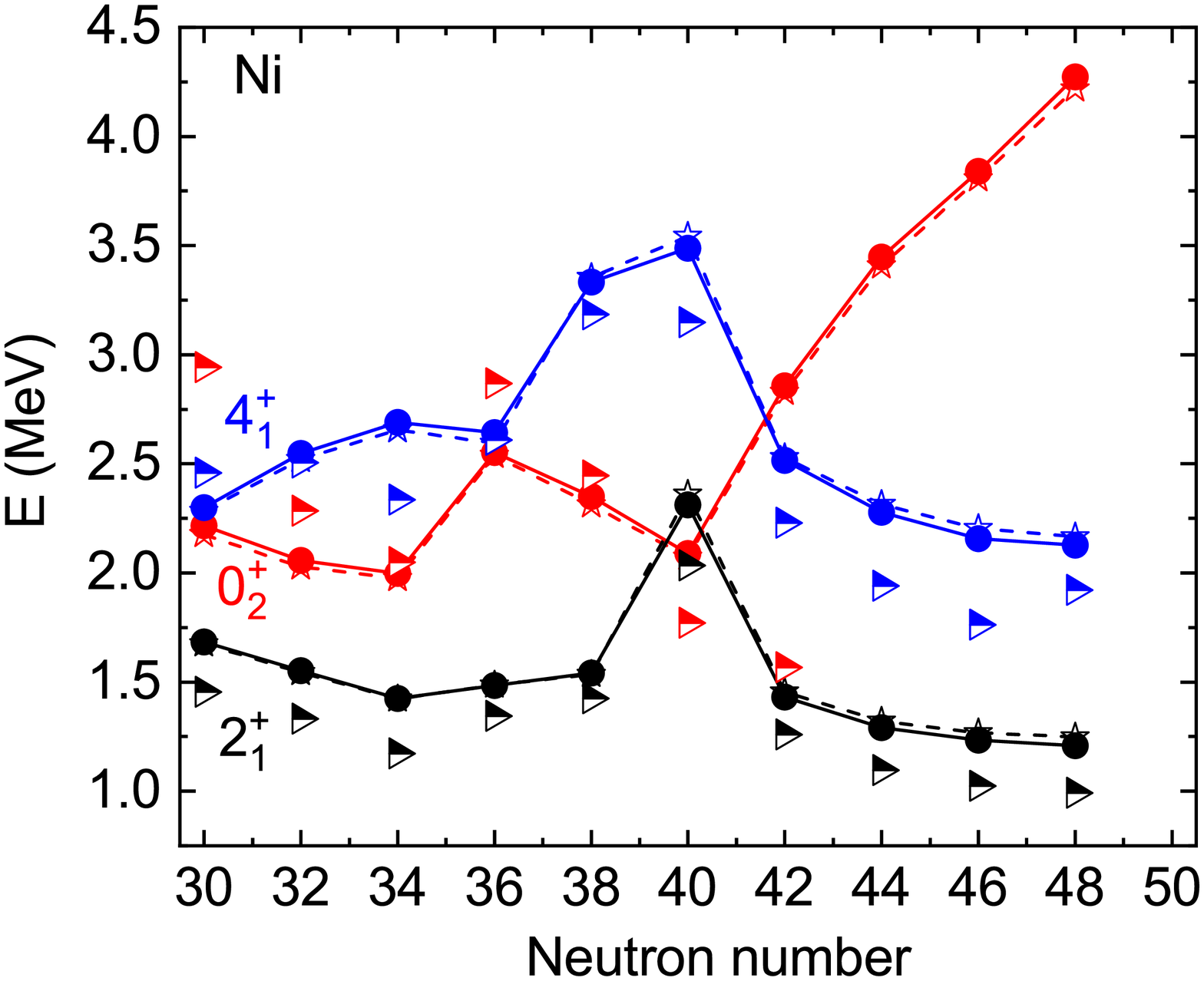}
	\end{subfigure}
	\begin{subfigure}{0.3\linewidth}
		\centering
		\includegraphics[width=1.30\textwidth]{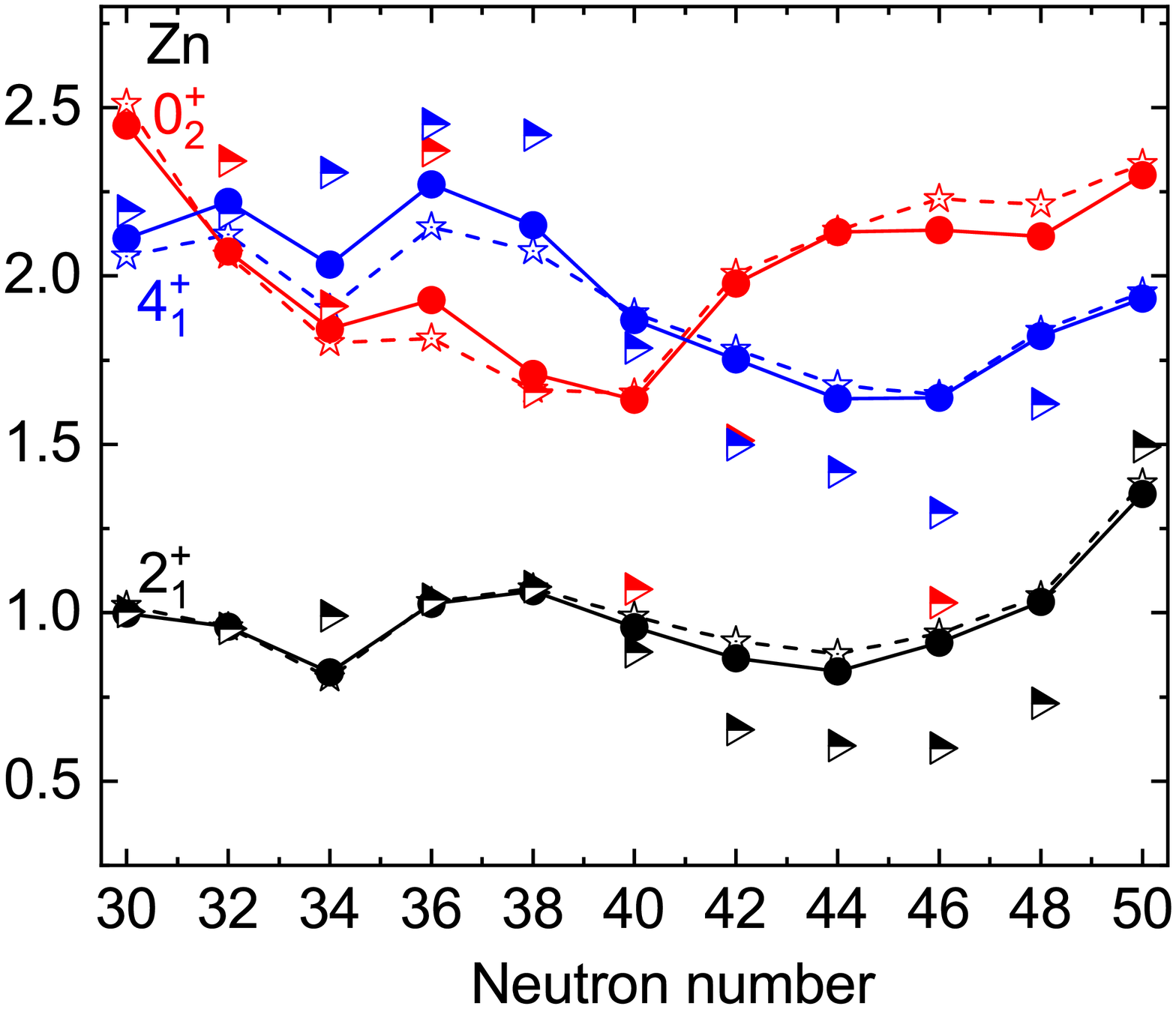}
	\end{subfigure}
	\begin{subfigure}{0.3\linewidth}
		\centering
		\includegraphics[width=1.30\textwidth]{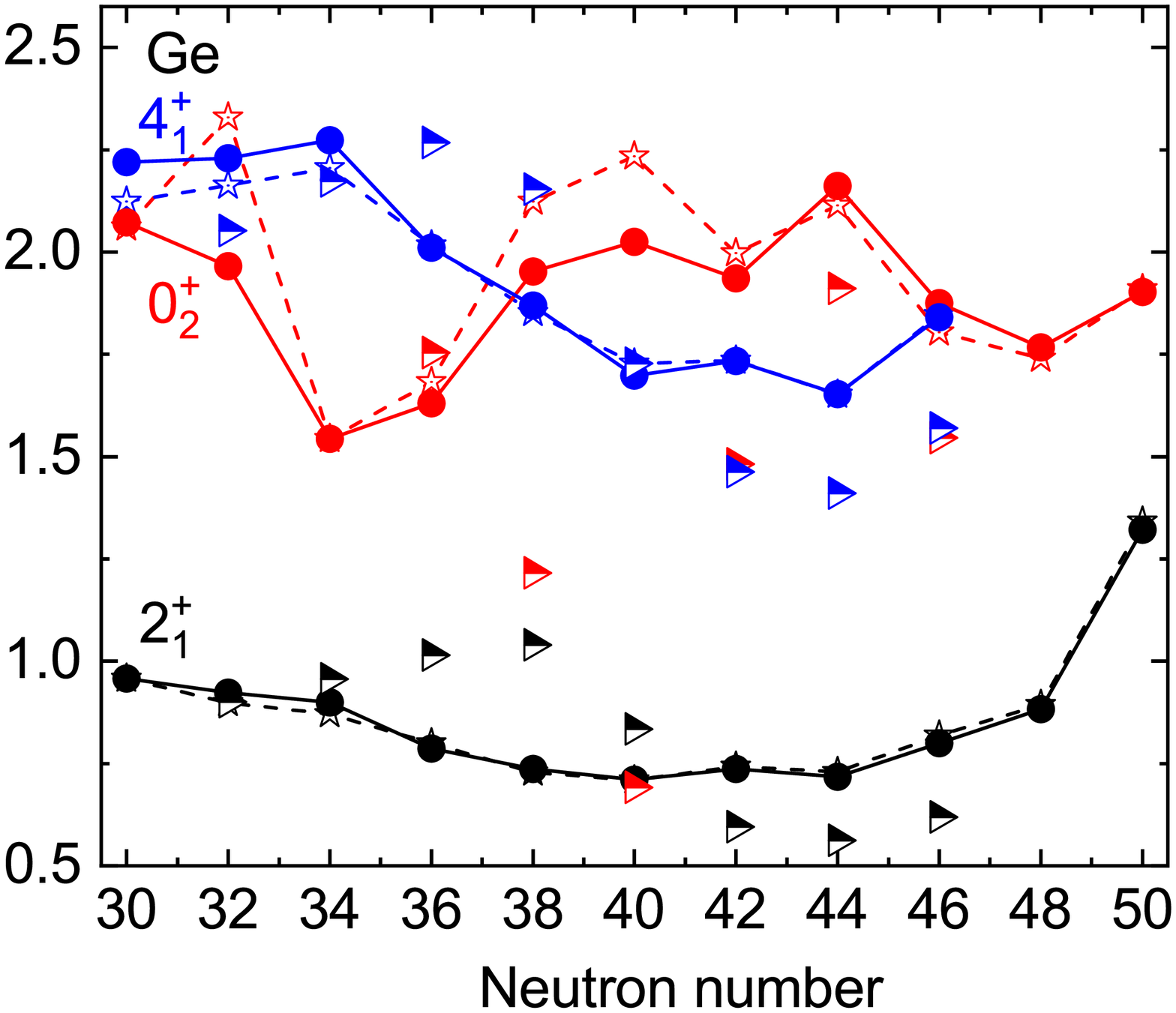}
	\end{subfigure}
	\vspace{-0.2cm}
	\caption{(Color online)  Low lying $2^{+}_{1}$, $4^{+}_{1}$ and $0^{+}_{2}$ of Ni, Zn, and Ge isotopes. Theoretical calculation are performed with the interactions jj44a (open star with dashed lines) and jj44b (solid circle with solid lines). The experimental data are shown with symbol half-filled triangle.}
	\label{F8}
\end{figure*}
\begin{figure*}
	\begin{subfigure}{0.3\linewidth}
		\centering
		\includegraphics[width=1.18\textwidth]{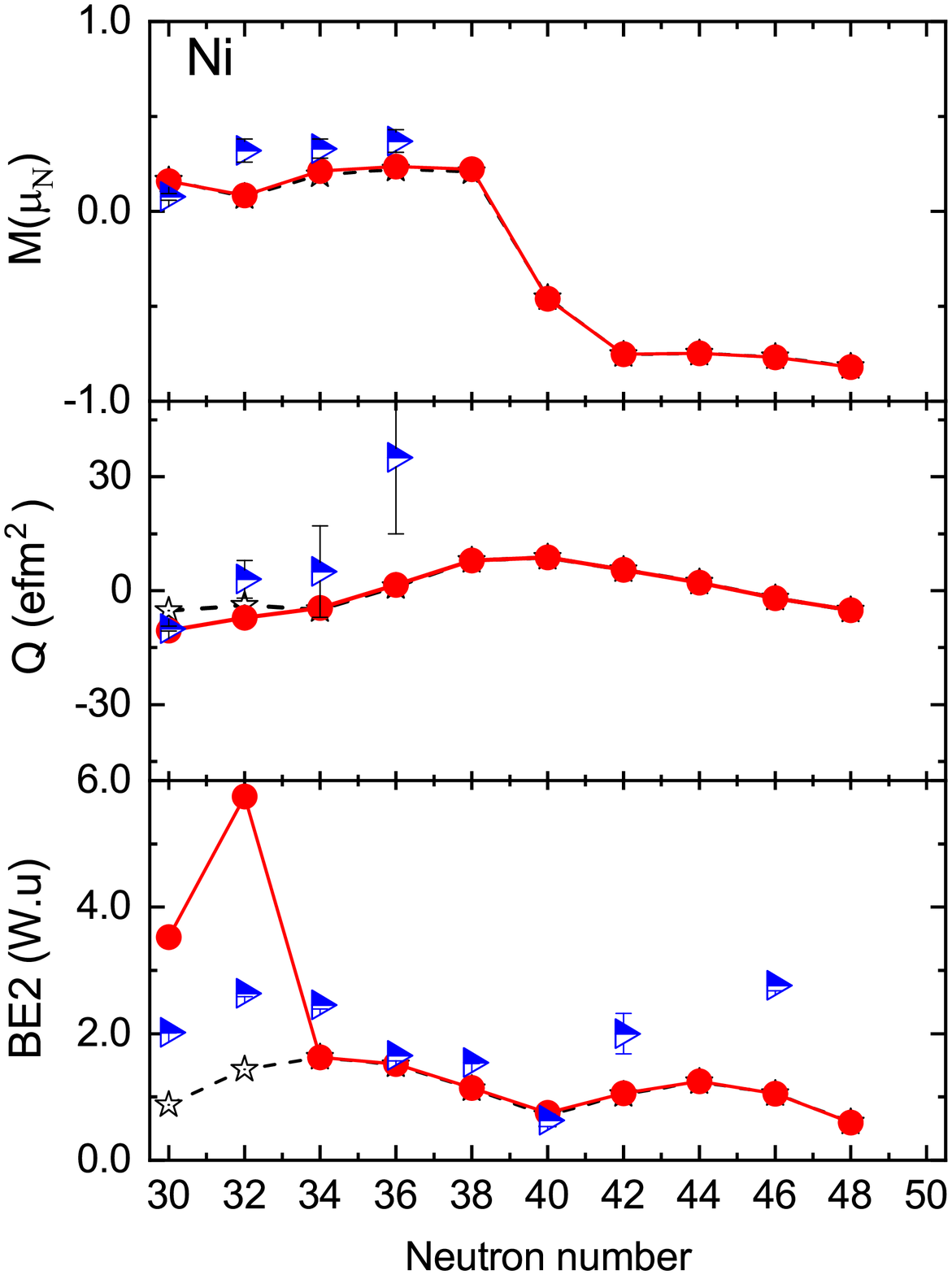}
	\end{subfigure}
	\begin{subfigure}{0.3\linewidth}
		\centering
		\includegraphics[width=1.18\textwidth]{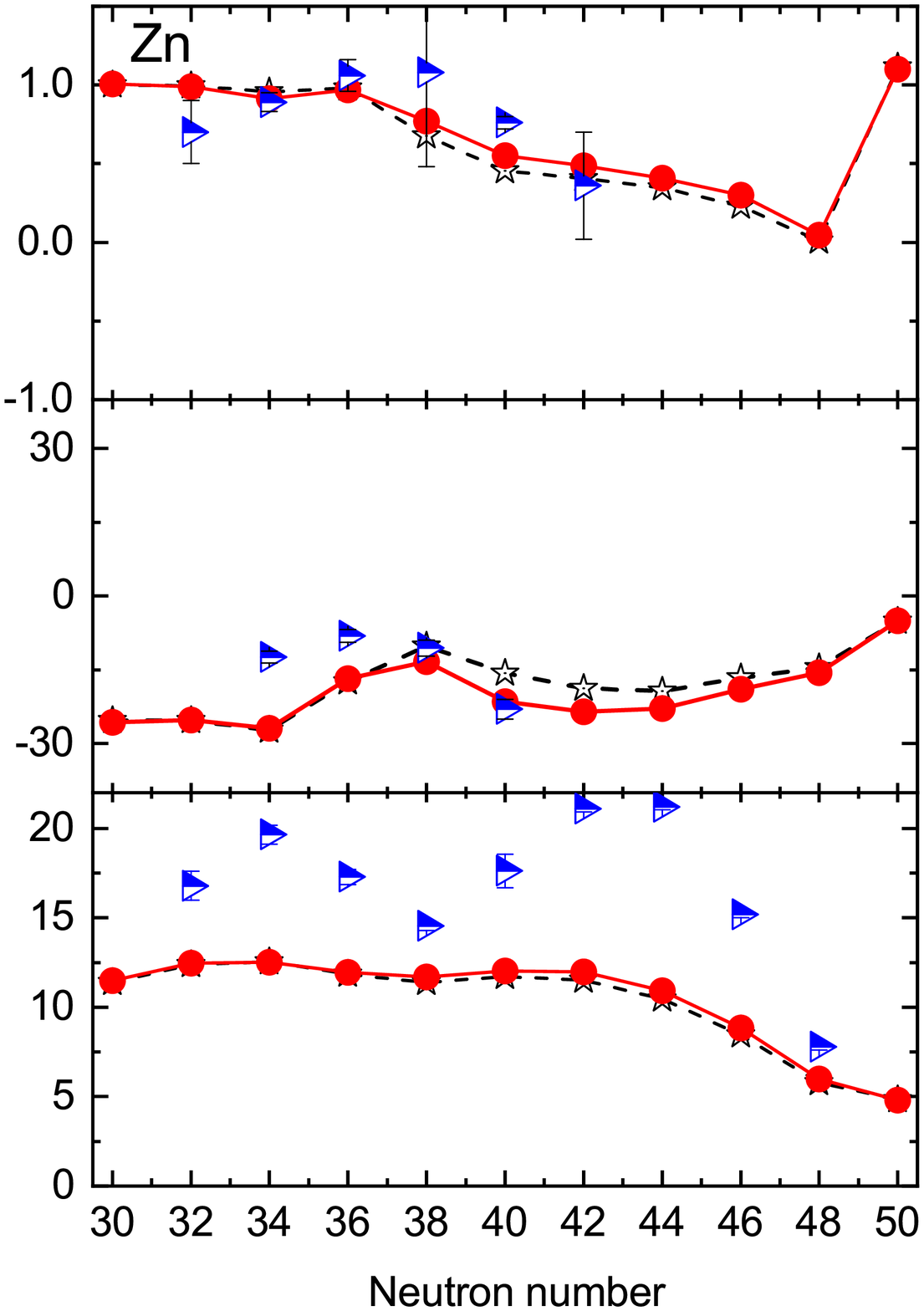}
	\end{subfigure}
	\begin{subfigure}{0.3\linewidth}
		\centering
		\includegraphics[width=1.18\textwidth]{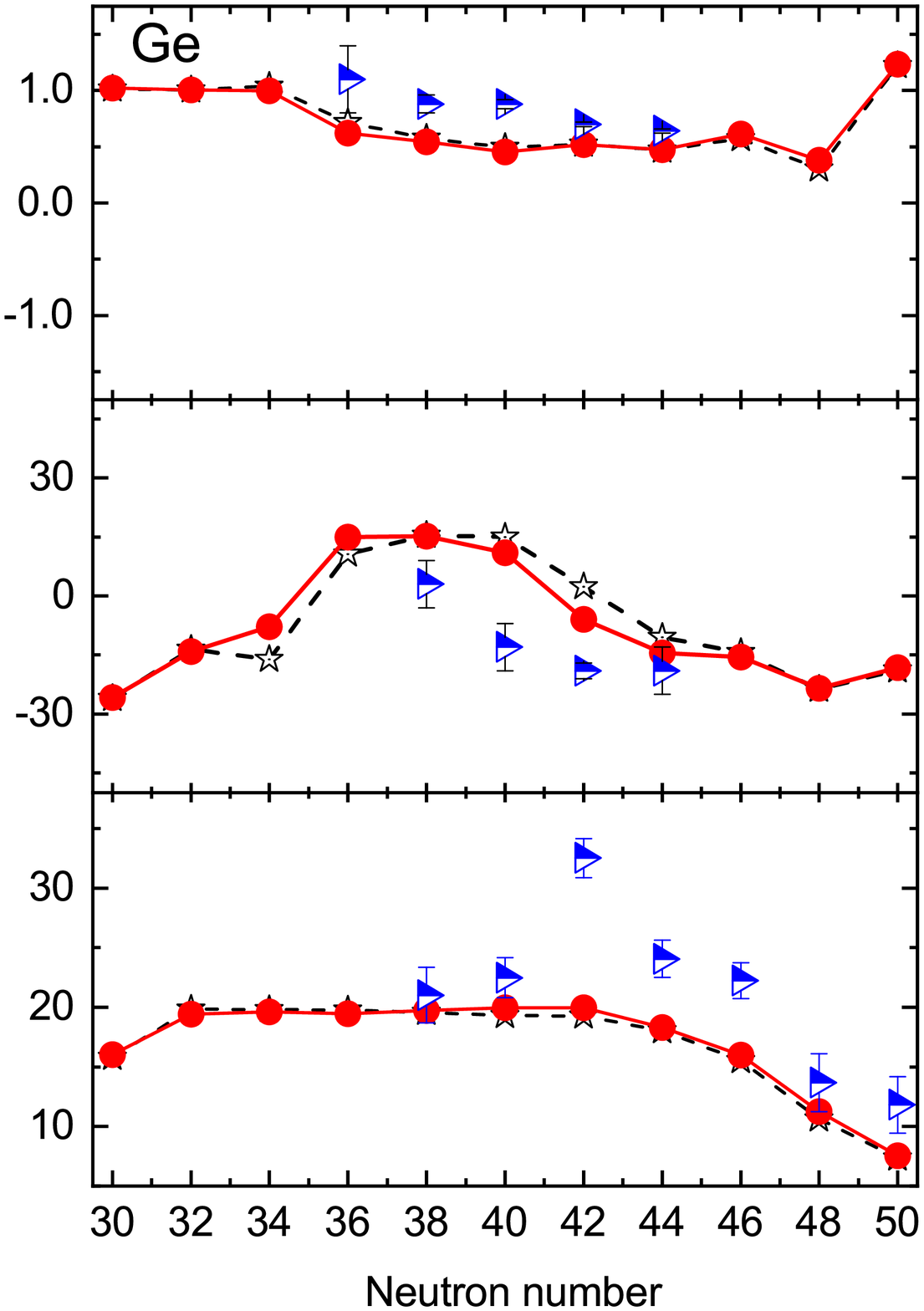}
	\end{subfigure}
\vspace{-0.5cm}
	\caption{(Color online) Magnetic moments (M), quadrupole moments (Q), and $B(E2; 2_{1}^{+} \rightarrow 0_{1}^{+})$ values for even-even Ni, Zn and Ge isotopes. The shell-model calculations are performed with interactions jj44a (open star with dashed lines) and jj44b (filled circle with solid lines) and experimental data (half-filled triangle) is taken from the Ref.\cite{nndc}}
	\label{F9}
\end{figure*}

\subsection{Electromagnetic moments and transitions}
In the previous section, we have investigated tensor force effects on the excitation energies and found that the tensor force has a significant role in the excitation spectra of Cu isotopes, particularly for the 5/2$_{1}^{-}$ state. The role of tensor force on lowering of 5/2$_{1}^{-}$ state with the filling of $\nu 0g_{9/2}$ orbital has been well recognized.  Now, in a similar way, we examine the tensor force effect on the electromagnetic observables as electromagnetic observables are a very important tool to probe the nuclear wavefunction and shape of the nuclei. Therefore, we have shown the magnetic moments and quadrupole moments of Cu isotopes with and without tensor force in Fig.~\ref{F5}. For present purposes, we have used free-nucleon $g$ factors $g_{s_{p}}$ = 5.586,  $g_{l_{p}}$= 1, $g_{s_{n}}$ = -3.826 and $g_{l_{n}}$ = 0, and standard effective charges $e_{p}$ = 1.5 e and $e_{n}$ = 0.5 e. The theoretically calculated magnetic moments of the 3/2$^{-}$ state of Cu isotopes are predicted higher than the experiment, however, the trend is remarkably consistent compared to the measured value. The magnetic moments are overestimated in the calculation due to the missing proton excitation across Z = 28 \cite{vingerhoets}. The interaction jj44b predicts peak at N = 40 for the magnetic moments of 3/2$^{-}$ state, which is similar to the experimental one. Further, the reported trends of ground state magnetic moments and spin of $^{75}$Cu is reasonably reproduced by the interaction jj44b \cite{flanagan, ichikawa}.  In Fig.~\ref{F5}, the sudden change in magnetic moments of $^{75}$Cu is indicative of structural change due to the change in spin \cite{flanagan}. It can be seen from Fig.~\ref{F5} that the tensor force has a small effect on the calculated  $\mu (3/2^{-})$ whereas the significant contribution for $\mu (5/2^{-})$.  

The theoretical calculation of quadrupole moments of Cu isotopes using interactions jun45 and jj44b are already discussed in the Ref.\cite{vingerhoets}. In this study, the effective charges $e_{p}$ = 1.5 e and $e_{n}$ = 1.1 e has been used to obtain the best result with respect to the experimental data. We have used the standard value of effective charges in the present work to explore the tensor force. Although, in the present work, the interaction jj44b overestimates the experimental data, however,  reproduces the trends very well. The calculated quadrupole moments at N = 40 is similar to the measured value within error bars,  but, overestimated beyond N = 40 \cite{flanagan}. This indicates the need for including proton excitations across Z = 28 \cite{sieja}. As from Fig.~\ref{F5}, the tensor force has a negligible effect on the ground state spin $(3/2^{-})$ whereas the significant contribution for spin $(5/2^{-})$.

In lower panel of Fig.~\ref{F6}, the transition probabilities of B($E2; 1/2^{-} \rightarrow 3/2^{-}$), B($E2;$ $ 5/2^{-} \rightarrow 3/2^{-}$) and B($E2; 7/2^{-} \rightarrow 3/2^{-}$) are shown using interactions jj44a and jj44b  for $^{69-77}$Cu. The calculated B(E2) follows similar trends as reported in the literature \cite{srivastava,stefanescu}. The excitation energies of the corresponding transitions are also shown in the upper panel of Fig.~\ref{F6}. The small value ($<$ 5 W.u.) of B($E2;5/2^{-} $ $ \rightarrow 3/2^{-}$) for $^{69-73}$Cu indicate that the 5/2$^{-}$ level possess single particle nature to  adequate extent. The small B($E2;5/2^{-} $ $ \rightarrow 3/2^{-}$) for $^{69}$Cu, supports the N = 40 magic shell gap. On the other hand large ($>$ 5 W.u.) B($E2; 1/2^{-} \rightarrow 3/2^{-}$) transition indicates that the 1/2$^{-}$ state depart from the single particle nature and develop collectivity. The calculation shows lowering of B($E2; 7/2^{-} \rightarrow $ $ 3/2^{-}$) transition from $^{63}$Cu and reach minimum at N = 40 whereas it is measured at N = 38 in the experiment. It can be seen from Fig.~\ref{F6} that the tensor force has significant contribution in excitation spectra, in particular 5/2$^{-}$ state whereas only small change can be noticed in B($E2$) value due to the tensor force. 

\subsection{Occupation number and spectroscopic factor}
To get more theoretical insight into the structure, we have shown the proton occupation numbers of proton orbitals in the upper part of the Fig.~\ref{F7}. The trends of the protons orbitals are very interesting. The occupancies of $\pi 0f_{5/2}$ orbital increases smoothly and cross $\pi 1p_{3/2}$ orbital at N = 40. The occupancies of $\pi 1p_{1/2}$ orbital is almost half of $\pi 1p_{3/2}$ orbital up to N = 38 and beyond this the occupancies of $\pi 1p_{1/2}$ orbital decreases smoothly. It is clear from Fig.~\ref{F7} that the occupancies of $\pi 0f_{5/2}$ orbital in $5/2^{-}$ state increases as the neutrons added in the orbitals which indicates that the $5/2^{-}$ state in heavier Cu isotopes have adequate single-particle nature. Contrary to this, the occupancies of $\pi 1p_{1/2}$ orbital in $1/2^{-}$ state decreases which indicate that the significant collectivity is present in $1/2^{-}$ state of heavier Cu isotopes. Further, like the previously discussed results of excitation energies and electromagnetic moments, the tensor force very much effects the occupancy number of $0f_{5/2}$ orbital. The lower panel of Fig.~\ref{F7} shows
the theoretical spectroscopic strength (C$^{2}$S) of $3/2^{-}$, $5/2^{-}$, and $1/2^{-}$ states using interactions jj44a and jj44b. To assess the proton single-particle strength in $3/2^{-}$, $5/2^{-}$, and $1/2^{-}$ states of Cu isotopes, the C$^{2}$S of $^{A}$Cu isotopes have been calculated with respect to $^{A-1}$Ni isotopes. These states are associated with the transfer of a proton to Ni which may occupy either of $\pi 1p_{3/2}$, $\pi 0f_{5/2}$,  and $\pi 1p_{1/2}$ orbitals, respectively. It can be followed from the Fig.~\ref{F7} that the $3/2^{-}$ state of Cu isotopes has very good proton single-particle strength. The $1/2^{-}$ state carry about half of the proton single-particle strength for the light isotopes and further reduces when $\nu 0g_{9/2}$ orbital start filling with neutrons.  The $5/2^{-}$ state shows much more mixing for $^{59, 69}$Cu, but gains good proton single-particle strength from $^{69}$Cu onwards. As expected, the tensor force largely influences the single-particle strength of $5/2^{-}$ state. Further, the spectroscopic strength and occupation number of $1/2^{-}$ state shows almost similar trends, whereas different in case of $3/2^{-}$ and $5/2^{-}$ states, particularly at N = 40. 	

\subsection{Low-lying excitation spectra and electromagnetic properties of even-even nuclei}
In Fig.~\ref{F8}, the shell-model calculation of $2^{+}_{1}$, $4^{+}_{1}$ and $0^{+}_{2}$ states of Ni, Zn, and Ge isotopes are shown with the interactions jj44a and jj44b along with the available experimental data. The low lying states predicted by jj44b are in good correspondence with the experiment for the lower mass region, whereas, states are predicted higher than the measured values beyond N = 40. These deviations are arising possibly due to missing $2d_{5/2}$ and $1f_{7/2}$ orbitals in the present model space. For Ni isotopes having no protons in their valence orbitals, the T = 1 two-body matrix elements are important. Since T = 1 tensor force monopole matrix elements of interactions jj44a and jj44b are almost similar except $\bar{V}_{p_{1}p_{1}}^{T = 1}$, therefore, the tensor force effect is not much visible in case of low lying excitation spectra of Ni isotopes. However, the tensor force contribution is clearly noticeable in the case of Zn and Ge isotopes. The irregular behaviour of $0^{+}_{2}$ state has been reasonably described by the interaction jj44b. The $0^{+}_{2}$ changes drastically and reaches a minimum at N = 40. For Ni and Ge, the $0^{+}_{2}$ is measured below $2^{+}_{1}$ at N = 40 whereas above in case of Zn. The large deviations between the calculated and measured $0^{+}_{2}$ value using interaction jj44b are observed around N $=$ 40 in particular for $^{72}$Ge. As from Fig.~\ref{F8}, it is evident that the tensor force has minimum contribution in $0^{+}_{2}$ state of Ni isotopic chain whereas maximum in case of Ge isotopes.  

The shell-model results of magnetic moments and quadrupole moments of  Ni, Zn and Ge isotopes are shown in Fig.~\ref{F9} with interactions jj44a and jj44b. In the calculation, the above discussed standard value of free-nucleon $g$ factors and effective charges are used. The obtained results are reasonably good with the experimental data. For N $>$ 40, the lowering of magnetic moments in Ni and Zn isotopes are observed. The tensor force has almost negligible contribution in magnetic moments of Ni isotopes whereas small changed has been noticed in case of Zn isotopes beyond N = 40. The sudden change in magnetic moments at N = 48 is indicative of the structural change in case of Zn and Ge isotopes. The theoretical calculation of quadrupole moments along with the experimental data of Ni, Zn and Ge isotopes are shown in the Fig.~\ref{F9}. For N $>$ 40, the quadrupole moments of Ni isotopes decreases and changes the sign at N = 46. Likewise the magnetic moments of Ni isotopes, the tensor force has no role in quadrupole moments for N $>$ 40. On the other hand, the tensor force has significant contribution in quadrupole moments calculation of Zn and Ge isotopes for N $>$ 40. The sign of the quadrupole moment is predicted positive for $^{72}$Ge, whereas it comes negative in the experiment.

The B($E2; 2^{+} \rightarrow 0^{+}$) transition of even$-$even Ni, Zn and Ge isotopes are shown using interactions jj44a and jj44b in the lower panel of Fig.~\ref{F9}. The calculated B($E2; 2^{+} \rightarrow 0^{+}$) of Ni isotopes reaches a minimum at N = 40 and beyond this B(E2) enhances significantly. The high 2$^{+}$ excitation energy and low B(E2) at N = 40 in case of Ni isotopes suggesting the semi-magic structure at N = 40. It can be seen from Fig.~\ref{F9} that the tensor force is not much significant for B($E2$) value except at N = 30 and 32. In case of Zn, the 2$^{+}$ excitation energy and B(E2) value are calculated high and low at N = 38, in contrast to the Ni isotopes at N = 40. This indicates that the N = 40 subshell effect disappears for Zn isotopes. The calculated B(E2) value is too large but almost constant from N = 32 to 42 for Ge isotopes.  The B(E2) value for Ge isotopes is almost constant for 32 $<$ N $<$ 42, whereas sudden change for N $>$ 42 is indicative of structural change.

\section{Summary}
\label{sec4} 
In summary, the tensor force two-body matrix elements have been calculated using well known Yukawa potential for \textit{ pfg} model space. The method has been used in our recent study for \textit{p}, \textit{sd}, and \textit{pf} model space by making use of spin-tensor decomposition technique, and we find good agreement of calculated ones with the tensor force monopole matrix elements of widely used effective interactions. Here, in case of \textit{pfg} model space, we also find that the calculated tensor force monopole matrix elements possess the systematics features for both Isospin T = 0 and 1. The interactions with (jj44b) and without (jj44a) tensor part are then used to calculate effective single-particle energy, E(2$^{+}$) of Ni, Zn and Ge, the level structure of Cu isotopes and electromagnetic properties of Ni, Cu, Zn and Ge isotopes by the aim of exploring the tensor force effect on them. In our investigations, we find that excitation energy of $5/2^{-}$ state in heavier Cu isotopes drops significantly when tensor force is included to the interaction jj44a, and tensor force play an important role for reproducing the correct experimental ground state in $^{73}$Cu and gap between $5/2^{-}$ and $1/2^{-}$ in most of the cases. Although, the $1/2^{-}$ state is not strongly affected as the $5/2^{-}$ state, tensor force lowers its excitation energy throughout the Cu chain. Further, the spectroscopic factors and proton occupation numbers study of $^{59-79}$Cu indicate that the $5/2^{-}$ state is largely affected by tensor force followed by $1/2^{-}$ state. We have also investigated even-even Ni, Zn and Ge isotopes, and found large deviation in $4_{1}^{+}$ and $0_{2}^{+}$ states of Zn and Ge isotopes due to the tensor force. As expected, the change by tensor force in magnetic moments are more pronounced for spin $5/2^{-}$ of Cu isotopes, whereas negligible change is found for Ni, Zn and Ge isotopes. On the other hand, the quadrupole moments of Zn and Ge isotopes are improved by including tensor force. In the present work, though we did not exactly quantify the tensor force, its effect can be seen qualitatively by observing changes in the results. 

\section*{Acknowledgments}
K. Jha acknowledges the Ministry of Human Resource and Development, Government of India for providing financial support.

\end{document}